\documentclass[10pt,aps,twocolumn,nolinenumbers, superscriptaddress, notitlepage]{revtex4-2}
\usepackage{amsbsy}
\usepackage{amsmath,amssymb}
\usepackage[usenames]{color}
\usepackage[colorlinks,linkcolor=black,citecolor=black]{hyperref}
\usepackage{graphicx}
\usepackage{xspace}
\usepackage{enumitem}
\usepackage{xcolor}
\usepackage{bm}
\usepackage{soul}
\usepackage{subfigure}
\usepackage{times}
\usepackage{physics}
\usepackage{qcircuit}
\usepackage{supertabular}
\usepackage{tabularx}
\usepackage{soul}
\usepackage{multirow}
\usepackage{comment}
\usepackage{textpos}
\usepackage[skins]{tcolorbox}
\usepackage{xr-hyper}
\usepackage{tikz}
\usetikzlibrary{positioning}

\definecolor{dbc}{rgb}{0.,0.4,.8}
\definecolor{ogc}{rgb}{0.4,.6,.5}

\newcommand{\cqt}{Centre for Quantum Technologies, National University of Singapore, Singapore 117543}
\newcommand{\maju}{MajuLab, International Joint Research Unit UMI 3654, CNRS, Université Côte d’Azur, Sorbonne Université, National University of Singapore, Nanyang Technological University, Singapore}
\newcommand{\cquere}{Center for Quantum Engineering Research and Education, TCG CREST, Techna, Sector V, Kolkata 700091, India}
\newcommand{\BSC}{Barcelona Supercomputing Center, Barcelona 08034, Spain}
\newcommand{\ICCUB}{Departament de F\'isica Qu\`antica i Astrof\'isica and Institut de Ci\`encies del Cosmos, Universitat de Barcelona, Barcelona 08028, Spain}
\newcommand{\Qilimanjaro}{Qilimanjaro Quantum Tech, Barcelona 08007, Spain}
\newcommand{\TII}{Quantum Research Centre, Technology Innovation Institute, P.O.Box: 9639, Abu Dhabi, United Arab Emirates}

\begin{document}

\title{Single-qubit universal classifier implemented on an ion-trap quantum device}

\author{Tarun Dutta}
\email{cqttaru@nus.edu.sg}
\address{\cqt}

\author{Adrián Pérez-Salinas}
\email{adrian.perezsalinas95@gmail.com}
\address{\BSC}
\address{\ICCUB}

\author{Jasper Phua Sing Cheng}
\address{\cqt}

\author{José Ignacio Latorre}
\address{\cqt}
\address{\Qilimanjaro}
\address{\TII}

\author{Manas Mukherjee}
\email{manas.mukh@gmail.com}
\address{\cqt}
\address{\maju}
\address{\cquere}

\begin{abstract}

Quantum computers can provide solutions to classically intractable problems under specific and adequate conditions. However, current devices have only limited computational resources, and an effort is made to develop useful quantum algorithms under these circumstances. This work experimentally demonstrates that a single-qubit device can host a universal classifier. The quantum processor used in this work is based on ion traps, providing highly accurate control on small systems. The algorithm chosen is the re-uploading scheme, which can address general learning tasks. Ion traps suit the needs of accurate control required by re-uploading. In the experiment here presented, a set of non-trivial classification tasks are successfully carried. The training procedure is performed in two steps combining simulation and experiment. Final results are benchmarked against exact simulations of the same method and also classical algorithms, showing a competitive performance of the ion-trap quantum classifier. This work constitutes the first experimental implementation of a classification algorithm based on the re-uploading scheme.

\end{abstract}

\maketitle

\section{Introduction}\label{sec:intro}

Quantum computing is an active field of research both in academia and industry. It is expected that quantum computers will provide solutions to classically intractable problems. A handful of problems has been found for which quantum computers can provide a solution with a certain, even exponential, speed-ups with respect to a classical computer \cite{Bern1997,Shor, Van2001, Deb2016,Mas2021}. On the experimental side, recent advances have achieved quantum supremacy, that is, solving a problem in a quantum computer more efficiently than using classical methods. This has been demonstrated both in superconducting~\cite{Arute_quantum_2019} and photonic~\cite{Zuchongzhi_2020, Zuchongzhi2_2021} platforms.  

Finding quantum supremacy for practical problem is an open question. Among various possibilities, quantum machine learning is considered to be a potential game changer. Data classification is an ubiquitous problem appearing across many different fields. Classical machine learning provides a plethora of algorithms for this purpose, notably those related to support vector machines~\cite{cortes_svm_1995} or neural networks~\cite{saad_nn_1999, doi:10.1063/1.1144830}. In recent years, classical machine learning has improved substantially and can now provide solutions to a large variety of classification problems~\cite{russell_artificial_2010, mohri_foundations_2012}. These classical algorithms are generally implemented using specialized software and hardware that are expensive in both computational and energy resources. 

Quantum Machine Learning (QML) aims to combine the spirits of classical machine learning and the capabilities of quantum computing. This field of research is still in an early stage and, in general, cannot compete against the state-of-the-art classical algorithms. However, recent theoretical works have proven quantum advantages in selected problems~\cite{liu_rigorous_2020}. Nowadays, most QML algorithms rely on variational methods~\cite{dunjko_machine_2018, schuld_circuit_2020, biamonte_quantum_2017,reviewvariational_2021}, where theoretical arguments assessing robust and general quantum advantages of quantum devices are still missing. Variational quantum algorithms are believed to fit the requirements of Noisy Intermediate-Scale Quantum~(NISQ) devices~\cite{Preskill2018quantumcomputingin}, motivating the use of these methods for the early experiments of QML. These experiments 
have progressed in different platforms \cite{QSVM_exp, QSVM_proposal, havlek_supervised_2019}. Photonic devices achieved one of the earliest experiments for addressing a classification problem~\cite{QSVM_exp} based on Quantum Support Vector Machines \cite{QSVM_proposal}.  
Superconducting qubits have addressed QML using variational algorithms~\cite{havlek_supervised_2019}.%

In this paper we implement a QML supervised learning algorithm on an ion-trap quantum device. This device is referred throughout this paper as a Quantum Processing Unit~(QPU). Trapped ions are an alternative design of quantum devices whose main strengths are the accurate control of small quantum systems~\cite{chemistry_ions,resch_quantum_2019}.  Recent experiments have widened the range of feasible experiments~\cite{figgatt_complete_2017, Nam_groundstate_2020, johri_nearest_2020, rudolph_generation_2020, PhysRevA.84.030303, Wang2021}. The QML algorithm chosen in this paper for this application is {\sl data re-uploading}~\cite{perez-salinas_one_2021, perez-salinas_data_2020}. The main reason for this choice is that data re-uploading is a method specifically designed to take advantage of the fixed Hilbert space available in quantum systems with few qubits. Thus, in order to attain good performances it is necessary to execute highly accurate quantum operations. Ion traps perfectly match these requirements.
The data re-uploading algorithm is a general scheme for supervised learning in QML using classical data. It was originally conceived as a quantum classifier~\cite{perez-salinas_data_2020}. The differentiating characteristic of this method is that the data is re-uploaded several times along the computation process. The performance of this algorithm improves as the query complexity, that is the number of data re-uploadings, increases. The theoretical capability of this model to approximate any function is guaranteed and emerges from its quantum nature~\cite{perez-salinas_one_2021, schuld_effect_2021}. 

The problem addressed in this paper is a supervised learning classification of synthetic data corresponding to geometrical figures as in Ref.~\cite{perez-salinas_data_2020}. The data re-uploading scheme for classification was already tested on classical simulators with noise environments~\cite{easom_depth_2021}. However, a classical simulator does not necessarily capture the merits and demerits of a real QPU. Indeed, a classical simulator is agnostic to the QPU platform specificity and it does not comprise implementation details such as native gates and specific noise models. The experimental approach presented in this work is, to the best of our knowledge, the first successful experimental implementation of the re-uploading scheme performing quantum classification with a single-qubit quantum processor. On the other hand, the present work accomplishes and surpasses experiments on superconducting qubits for the single-qubit approximant for function regression \cite{perez-salinas_one_2021}.

The paper is structured as follows: Sec. \ref{sec:framework} explains the experiment including both theoretical and implementation aspects. Sec.~\ref{sec:results} details the performance of the classifier in several examples. A discussion on the results is carried in Sec.~\ref{sec:discussion}. Some more information on experiment and results can be found in the Appendix.

\section{Framework}\label{sec:framework}
The problem solved in this work corresponds to supervised learning. In supervised learning, a model is fed with some data in the form $\mathbb{D}_{\rm train} = \{\vec x, c\}$, where $\vec x$ is a feature point and $c$ is the class it belongs to. The algorithm then learns the latent properties of the dataset in such a way that it can predict the class of previously unseen points belonging to another equivalent dataset $\mathbb{D}_{\rm test}$. 

The algorithm is executed on an ion-trap quantum machine, see Sec.~\ref{sec:experiment}. The method of data re-uploading is briefly explained in Sec.~\ref{sec:reuploading}. The optimization procedure is depicted in Sec.~\ref{sec:optimization}. Final experiment-dependents implementations to optimize performance are detailed in Sec.~\ref{sec:opt_hardware}.

\subsection{Experimental setup}\label{sec:experiment}
\begin{figure}[t!]
\definecolor{myBlue}{HTML}{9dc3e6}
\definecolor{myOrange}{HTML}{c65d14}
\definecolor{myGreen}{HTML}{00b843}
\subfigure[Ion-trap architecture]{\includegraphics[width=.48\linewidth]{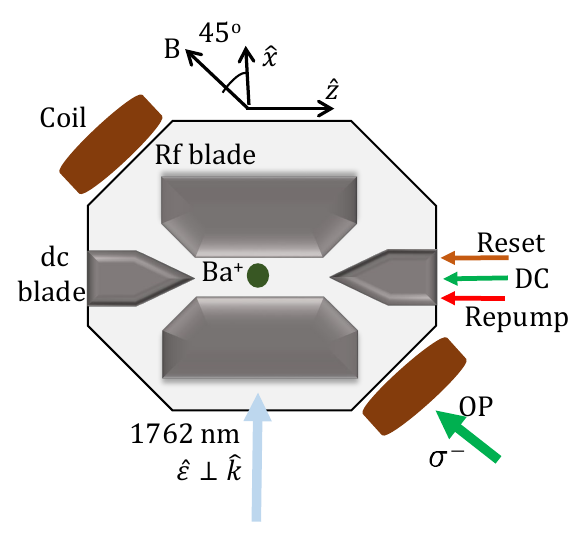}} 
\subfigure[Energy levels of the $^{138}$Ba$^+$ ion]{
\resizebox{.48\linewidth}{!}{
\begin{tikzpicture}

\draw[myBlue, very thick, <->] (1.625,-0.2) -- (4.625,4*.65 - .2);
\draw (3.8,.34) node {$1762$ nm};
\draw[myOrange, very thick, <->] (.55,2*2.59) -- (3.5,3*.91);
\draw (3,2 * 1.8) node {$614$ nm};
\draw[red, very thick, <->] (3.5,2*.61) -- (.6,2*2.05);
\draw (2.5,1.5*1.4) node {$650$ nm};
\draw[myGreen, very thick, <->] (.4,.05) -- (.4,2*2.05);
\draw (-.3, 2) node {$493$ nm};
\draw[myGreen, very thick, <->] (1.625,.2) -- (1.625,2*2);
\draw (1.4, 2*.95) node {$\sigma^{-}$ };

\draw[black, thick] (0,0)node[anchor=east]{$S_{1/2}$} -- (1.5,0) ;
\draw[black, thick] (0,2*2.1)node[anchor=east]{$P_{1/2}$} node[anchor=south west]{\scriptsize $\tau = 7.8$~ns} -- (1.5,2*2.1) ;
\draw[black, thick] (0,2*2.6)node[anchor=east]{$P_{3/2}$} node[anchor=south west]{\scriptsize $\tau = 6.3$~ns}-- (1.5,2*2.6);
\draw[black, thick] (3,2*0.60)node[anchor=east]{$D_{3/2}$} -- (4.5,2*0.60) ;
\draw[black, thick] (3,3*0.90)node[anchor=east]{$D_{5/2}$} node[anchor=south west]{\scriptsize $\tau = 32$~s} -- (4.5,3*0.90);

\draw[black] (1.5,0.2) -- (1.75,0.2) node[anchor=west]{\small $+1/2$};
\draw[black] (1.5,-0.2) -- (1.75,-.2) node[anchor=west]{\small $-1/2$};
\draw[black] (1.5,2*2) -- (1.75,2*2) node[anchor=west]{\small $+1/2$};
\draw[black] (1.5,2*2.2) -- (1.75,2*2.2) node[anchor=west]{\small $-1/2$};
\draw[black] (4.5,5*0.56 + .2) -- (4.75,5*0.56 + .2) node[anchor=west]{\small $+1/2$};
\draw[black] (4.5,4*0.64 - .2) -- (4.75,4*0.64 - .2) node[anchor=west]{\small $-1/2$};
\draw[black] (4.5,5*0.56 + .5) -- (4.75,5*0.56 + .5) node[anchor=west]{\small $+3/2$};
\draw[black] (4.5,4*0.64 - .5) -- (4.75,4*0.64 - .5) node[anchor=west]{\small $-3/2$};
\draw[black] (4.5,5*0.56 + .8) -- (4.75,5*0.56 + .8) node[anchor=west]{\small $+5/2$};
\draw[black] (4.5,4*0.64 - .8) -- (4.75,4*0.64 - .8) node[anchor=west]{\small $-5/2$};

\end{tikzpicture}
}
}
\caption{a) Schematic drawing of the ion-trap setup viewed from a top showing trap orientation and laser beam directions. The $^{138}$Ba$^+$ ion is confined in a Linear Paul trap. Two coils in Helmholtz configuration generate a magnetic field (B) defining the quantization axes.  b) Energy level diagram of the ion showing the Zeeman splitting of atomic states $S_{1/2}$, $P_{1/2}$ and $D_{5/2}$. The degeneracy of the Zeeman sublevels is lifted by a magnetic field of $1.5$ G. Laser light at 493 nm is used for Doppler-cooling, optical pumping and detection. The lasers at 650 nm and 614 nm pump out the D-states. An ultra-stable and narrow linewidth laser at 1762 nm is used for manipulating the qubit encoded in the quadrupole transition.}

\label{fig:trap_energy}
\end{figure}
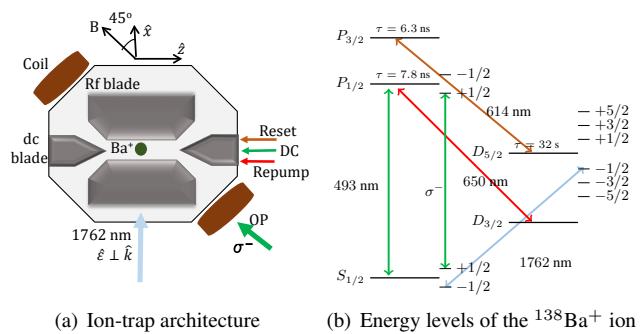

The qubit is realized in a $^{138}$Ba$^+$ trapped ion. The computational basis corresponds to the states 
\begin{equation}
    \ket 0 \equiv~{\rm S}_{\frac{1}{2},-\frac{1}{2}} \qquad \ket 1 \equiv~{\rm D }_{\frac{5}{2},-\frac{1}{2}}.
\end{equation}
 Both states are coupled by the electric quadrupole E$2$ transition at $1762~$nm wavelength shown in Fig.~\ref{fig:trap_energy}(b). The ion is confined in a linear blade Paul trap, as shown in Fig.~\ref{fig:trap_energy}(a), operating at a radial frequency of $1.5~$MHz and an axial frequency of $0.3~$MHz. The most relevant parameters of the trapped ion in this experiment are the qubit coherence time~($5~$ms) and the Rabi $\pi$-time of the qubit~($12~\mu$s)~\cite{Yum:17,Dutta2020}. The qubit is well-characterized in terms of both its internal~\cite{Dutta2020} and external degrees of freedom~\cite{VanHorne2020}.

Prior to performing each algorithmic cycle, the qubit is initialized to the state $\text{S}_{\frac{1}{2},-\frac{1}{2}}$. To do so, the qubit is Doppler-cooled to the Lamb-Dicke regime~\cite{Dutta2020} via a fast dipole transition between S-P levels at $493~$nm and a re-pump laser between D-P levels at $650~$nm shown in the Fig.~\ref{fig:trap_energy}(b). In this regime the internal and external states of the qubit are decoupled and hence single qubit gate errors are not influenced by the motion of the qubit. Then, the alternative ground state S$_{\frac{1}{2},+\frac{1}{2}}$ is selectively de-populated by optical pumping using a $\sigma^-$ polarized 493 nm laser together with an 650 nm laser pulse~\cite{Dehmelt1957}. This propagates along the quantization axis defined by the direction of an externally applied magnetic field $B$. The strength and direction of the $B$ field is optimized to achieve high fidelity state initialization $\ge 99\%$ within an optical pumping time $\le 50\mu$s. The magnetic field strength is set at $1.5$G, directed at an angle of $45^{o}$ with respect to the trap axis (z-axis) as illustrated in Fig.~\ref{fig:trap_energy}(a). The resultant strength of the magnetic field corresponds to a ground state Zeeman splitting of about $2~$MHz.
Once the qubit is initialized, any single qubit rotational gate is implemented by resonantly driving the qubit with full control over the laser phase, power and laser on-time. The qubit is operated by means of an ultra-low linewidth laser operated at $1762~$nm. The laser is phase locked to a ultra-low expansion cavity achieving an linewidth $\le 100~$Hz. This linewidth is estimated from the measured atomic resonance. The laser-qubit interaction time sets the rotation angle. It is controlled by a radio-frequency~(RF) switch that implements pulse-shaping thus reducing off-resonant phase noise in the circuit. An acousto-optic modulator (AOM) controls the phase and frequency of the laser implementing the rotation gates. Therefore, direct and precise control over axis and angle of rotations is achieved. 

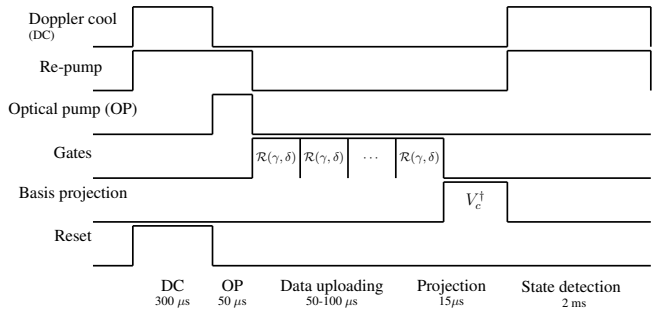
\begin{figure}[t!]
\begin{center}
\resizebox{\linewidth}{!}{
\begin{tikzpicture}
\draw (0,0.5) node  [align=left] {\large Doppler cool\\(DC)};
\draw[very thick,-] (0.5,0) -- (1.5,0);
\draw[very thick,-] (1.5,0) -- (1.5,1);
\draw[very thick,-] (1.5,1) -- (3.5,1);
\draw[very thick,-] (3.5,1) -- (3.5,0);
\draw[very thick,-] (3.5,0) -- (10.9,0);
\draw[very thick,-] (10.9,0) -- (10.9,1);
\draw[very thick,-] (10.9,1) -- (14.5,1);
\draw[very thick,-] (14.5,1) -- (14.5,0);

\draw (0,-0.55) node  [align=left] {\large Re-pump };
\draw[very thick,-] (0.5,-1.1) -- (1.5,-1.1);
\draw[very thick,-] (1.5,-1.1) -- (1.5,-0.1);
\draw[very thick,-] (1.5,-0.1) -- (4.5,-0.1);
\draw[very thick,-] (4.5,-0.1) -- (4.5,-1.1);
\draw[very thick,-] (4.5,-1.1) -- (10.9,-1.1);
\draw[very thick,-] (10.9,-1.1) -- (10.9,-0.1);
\draw[very thick,-] (10.9,-0.1) -- (14.5,-0.1);
\draw[very thick,-] (14.5,-0.1) -- (14.5,-1.1);

\draw (0,-1.6) node  [align=left] {\large Optical pump~(OP)};
\draw[very thick,-] (0.5,-2.2) -- (3.5,-2.2);
\draw[very thick,-] (3.5,-2.2) -- (3.5,-1.2);
\draw[very thick,-] (3.5,-1.2) -- (4.5,-1.2);
\draw[very thick,-] (4.5,-1.2) -- (4.5,-2.2);
\draw[very thick,-] (4.5,-2.2) -- (14.5,-2.2);

\draw (0.0,-2.65) node  [align=left] {\large Gates};
\draw[very thick,-] (0.5,-3.3) -- (4.5,-3.3);
\draw[very thick,-] (4.5,-3.3) -- (4.5,-2.3);
\draw[very thick,-] (4.5,-2.3) -- (9.3,-2.3);
\draw[very thick,-] (9.3,-2.3) -- (9.3,-3.3);
\draw[very thick,-] (5.7,-3.3) -- (5.7,-2.3);
\draw[very thick,-] (6.9,-3.3) -- (6.9,-2.3);
\draw[very thick,-] (8.1,-3.3) -- (8.1,-2.3);
\draw[very thick,-] (9.3,-3.3) -- (14.5,-3.3);
\draw (5.1,-2.8) node  [align=center] { $\mathcal R(\gamma,\delta)$};
\draw (6.3,-2.8) node  [align=center] { $\mathcal R(\gamma,\delta)$};
\draw (7.5,-2.8) node  [align=center] {$\cdots$};
\draw (8.7,-2.8) node  [align=center] { $\mathcal R(\gamma,\delta)$};

\draw (0.0,-3.7) node  [align=left] {\large Basis projection};
\draw[very thick,-] (0.5,-4.4) -- (9.3,-4.4);
\draw[very thick,-] (9.3,-4.4) -- (9.3,-3.4);
\draw[very thick,-] (9.3,-3.4) -- (10.9,-3.4);
\draw[very thick,-] (10.9,-4.4) -- (10.9,-3.4);
\draw[very thick,-] (10.9,-4.4) -- (14.5,-4.4);
\draw (10.1,-3.85) node  [align=center] {\large $V_c^\dagger$};

\draw (0.0,-4.75) node  [align=left] {\large Reset};
\draw[very thick,-] (0.5,-5.5) -- (1.5,-5.5);
\draw[very thick,-] (1.5,-5.5) -- (1.5,-4.5);
\draw[very thick,-] (1.5,-4.5) -- (3.5,-4.5);
\draw[very thick,-] (3.5,-5.5) -- (3.5,-4.5);
\draw[very thick,-] (3.5,-5.5) -- (14.5,-5.5);

\draw (2.5,-6.2) node  [align=center] {\large DC\\ 300 $\mu$s};
\draw (4,-6.2) node  [align=center] {\large OP\\ 50 $\mu$s};
\draw (6.5,-6.2) node  [align=center] {\large Data uploading\\ 50-100 $\mu$s};
\draw (9.5,-6.2) node  [align=center] {\large Projection\\ 15$\mu$s};
\draw (12.5,-6.2) node  [align=center] {\large State detection\\ ~ 2 ms};
\end{tikzpicture}}

\caption{Time sequence used in the experiment to perform each classifier measurement. At the beginning of each run, the qubit is reset to the $\ket 0 \equiv \ket{S_{1/2, -1/2}}$ state via Doppler cooling and optical pump. Then, the different reuploading $\mathcal R(\gamma,\delta)$ gates are applied sequentially via laser pulses. The duration of each pulse depends on the value of its $\beta$ parameter. As a final step, the state is projected to be compared against a label state defined by $V_c\ket 0 = \ket{\phi_c}$ and the relative fidelity is measured.}
\label{fig:timeseq}
\end{center}
\end{figure}

A direct digital synthesizer (DDS) based on $AD9959$ chip is used to control the application of gates on the qubit. The DDS is disciplined by a rubidium frequency standard (\textit{SRS FS725}), eliminating the long term frequency drift of the DDS clock, thus maintaining the errors in phase relations of the sequential gates of the classifier below $0.01\%$. Since the parameters are uploaded to the circuit on-the-fly, the latency of uploading the DDS parameters play a crucial role. The latency is minimized by pre-loading the full sequence of the phase, frequency and power data to an on-chip memory of the DDS. The current version of the DDS controlling the phase of the laser is limited by the on-chip memory to $16$ phase modulation steps thus limiting the layers to six, which is sufficient for the current discussion. The DDS output is then controlled by an external trigger generated from a Field Programmable Gate Array~(FPGA) based pulse pattern generator that controls the time sequence of the entire experiment with a time jitter below $10~$ns. See Fig.~(\ref{fig:timeseq}) for a scheme of the experiment at the time-sequence level. 

The phase of the AOM is directly controlled by a DDS which supplies the RF signal to the AOM via an amplifier. In order to avoid accumulation of phase noise, during the on-off time of the laser, RF pulse shaping has been implemented at the RF switch before being fed to the AOM for switching. All the RF sources are synchronized to the DDS clock. The main upgrade with respect to our previous setup lies in the hardware to better control the qubit phase. In addition, we modified the control software to implement quantum algorithms requiring long circuit depth with low latency. 

For the measurement step, the final state is projected onto the computational state $\ket{0}$ and the ground state occupation probability is obtained by observing spontaneously emitted photons while the qubit is excited by the Doppler-cooling $493~$nm laser. With a photon collection time of $2~$ms, the qubit state is determined by choosing a photon-count threshold of $15~$counts/$2~$ms such that the bright state $\ket 0\equiv {\rm S}_{1/2}$ is clearly discriminated from the dark $\ket 1\equiv{\rm D}_{5/2}$ state. This projective measurement is $100\%$ efficient in discriminating both states. Taking repeated measurements on the same state provides the probability and hence the projection of the state along the $\sigma_z-$axis.

\subsection{Re-uploading scheme}~\label{sec:reuploading}

\begin{figure*}
\begin{flushleft}
\begin{tcolorbox}[enhanced,width=1\textwidth, center, colback=white]
\begin{center}

\Large \sl Quantum classifier: multi-variable $\vec x$ $\Rightarrow$ multi-class $\vec c$
\end{center}

\vskip5mm
\begin{flushleft}
{\bf 1. TRAINING USING CLASSICAL SIMULATION}\newline

\resizebox{.75\textwidth}{!}{\hskip1cm $\Qcircuit @R=0.5em @C=0.3em{
\lstick{\ket 0} & \qw & \gate{U(\vec x, \vec\theta^{\rm sim}_{1})} & \gate{U(\vec x, \vec\theta^{\rm sim}_{2})} & \qw & \push{\cdots} & \gate{U(\vec x, \vec\theta^{\rm sim}_{k})} &\qw & \meter & \ustick{\hspace{5mm} \Rightarrow}\cw & \control \cw\\
 &  & \control \cwx & \control \cwx\cw & \dstick{\Leftarrow}\cw & \cw & \control \cwx\cw & \gate{\rm Classical\; Optimizer} \cw & \dstick{\Leftarrow} \cw & \cw & \gate{\chi^2(\Theta^{\rm sim}, \mathbb{D}_{\rm train})} \cwx \\
}$}
\begin{textblock}{3}(9.4, -0.3)
{\Large$
\Rightarrow \Theta^{\rm sim}$
}
\end{textblock}
\begin{textblock}{3}(8.7, -0.7)
{$
\longrightarrow {\rm Prob} (\vec x, \Theta|\vec c)^{\rm \; sim}$
}
\end{textblock}
\vskip5mm

{\bf 2. TRAINING USING QUANTUM PROCESSING UNIT}\newline

\resizebox{.75\textwidth}{!}{\hskip1cm $\Qcircuit @R=0.5em @C=0.3em{
\lstick{\ket 0} & \qw & \gate{U(\vec x, \vec\theta^{q}_{1})} & \gate{U(\vec x, \vec\theta^{q}_{2})} & \qw & \push{\cdots} & \gate{U(\vec x, \vec\theta^{q}_{k})} &\qw & \meter & \ustick{\hspace{5mm} \Rightarrow}\cw & \control \cw\\
 & & \control \cwx & \control \cwx\cw & \dstick{\Leftarrow}\cw & \cw & \control \cwx\cw & \gate{\scriptsize \textrm{Experimental Optimizer}} \cw & \dstick{\Leftarrow} \cw & \cw & \gate{\mathcal{A}(\Theta^{q}, \mathbb{D}_{\rm test})} \cwx \\
}$}
\begin{textblock}{3}(9.4, -0.3)
{\Large
$\Rightarrow \Theta^{q} = \Theta^{\rm sim} + \delta\Theta$
}\end{textblock}
\begin{textblock}{3}(8.7, -0.7)
{$
\longrightarrow {\rm Prob} (\vec x, \Theta|\vec c)^{q}$
}
\end{textblock}
\end{flushleft}

\vskip2mm
\end{tcolorbox}
\caption{\label{fig:fig1} Schematic description of the optimization algorithm used in this work. 1) Data re-uploading is trained using a classical simulation. The simulated quantum circuit is composed of single-qubit gates $U$ that depend on variational parameters $\Theta = \lbrace \vec\theta_1, \vec\theta_2,\ldots,\vec\theta_L\rbrace$ and the variables $\vec x$ associated to a given pattern. The output state is measured to obtain a vector ${\rm Prob} (\vec x, \Theta|\vec c)$ encoding the probabilities that will serve to classify the given pattern into a category. A classical optimization is performed to obtain optimal values $\Theta^{\rm sim}$. This optimization is driven by the cost function $\chi^2$ evaluated on training data, $\mathbb D_{\rm train}$. 2) A further optimization is accomplished only using the quantum device, taking as starting point the values $\Theta^{\rm sim}$ and delivering a better set $\Theta^{q}$. The quantity which is now maximized is the accuracy $\mathcal A$ evaluated on the test dataset $\mathbb{D}_{\rm test}$. The aim of the experimental optimization is to mitigate and even compensate possible systematic experimental errors.}
\end{flushleft}

\end{figure*}
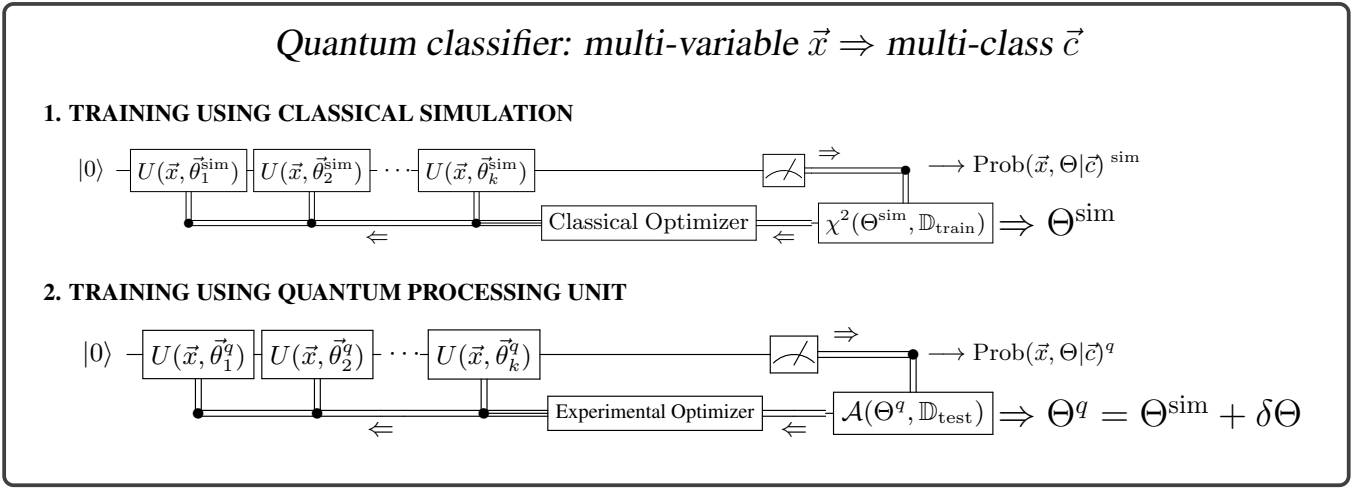

The re-uploading algorithm here presented delivers an output quantum state in the form
\begin{equation}\label{eq:gen_circuit}
    \ket{\psi(\vec x, \Theta)} = \prod_{i = 1}^L U(\vec{x}, \vec\theta_i)\ket 0
\end{equation}
where each $U(\vec{x}, \vec\theta_i)$ is referred to as a layer and depends on both sampling data $\vec x$ and tunable parameters $\vec{\theta_i}$, with $\Theta = \lbrace \vec\theta_1, \vec\theta_2,\ldots,\vec\theta_L\rbrace$. Each layer is composed by $R_y$ and $R_z$ rotations. In particular, two Ansätze are proposed in this work and inspired from Ref.~\cite{perez-salinas_data_2020}. Ansatz A holds for multidimensional data and introduces $\vec x$ only in $R_y$ rotations as
\begin{equation}\label{eq:ansatz_a}
U_A(\vec{x}, \vec\theta) =  R_z(\varphi) R_y(\vec{\omega} \cdot \vec x + \alpha). 
\end{equation}
On the other hand,
Ansatz B holds only for 2-dimensional datasets and introduces data in all rotation gates. It is defined as 
\begin{equation}\label{eq:ansatz_b}
U_B(\vec{x}, \vec\theta) =  R_z(\omega x_2 + \beta) R_y(\nu x_1 + \alpha).
\end{equation}
In both cases, $\vec\theta$ are flexible sets of parameters that accomodate the requirements of each gate. 

The next step consists in training the circuit to obtain a set of parameters $\Theta$ such that it solves a given problem. Since we aim to build a classifier for supervised learning, each sample $\vec x$ in the dataset is associated to a class within the set $\{\vec c\}$. We must also define as many label states $\ket{\phi_c}$ as existing classes. Then, the probability of a state corresponding to a class $c \in \vec c$ corresponds to 
\begin{equation}\label{eq:label_states}
{\rm Prob} (\vec x, \Theta|c) \propto \left\vert \braket{\phi_c}{\psi(\vec x, \Theta)}\right\vert^2,.
\end{equation}
properly normalized.
This probability assignment needs to choose a target ket $\ket{\phi_c}$ for each class in every different problem here considered. The general rule is to choose the target kets as distant as possible from each other, to maximize the distinguishability among the classes~\cite{perez-salinas_data_2020}. Given the output from the quantum computer, a final class $c$ is assigned to a given pattern according to the label ket that maximizes the probability 
${\rm Prob} (\vec x, \Theta|c)$. 

For the algorithm to work properly it is necessary to find the configuration of parameters $\Theta$ such that the  fidelity between the output states and their corresponding label states are in average maximum for all patterns in the training dataset, $\{\vec x, c\}\in \mathbb D_{\rm train}$. For a training dataset with $N_{\rm train}$ patterns, the loss function describing this behavior is just

\begin{multline}\label{eq:loss}
    \chi^2(\Theta;\mathbb{D}_{\rm train} ) = \\  \frac{1}{N_{\rm train}} \sum_{(\vec x, c) \in \mathbb D_{\rm train}}\left( \left\vert \braket{\phi_c}{\psi(\vec x, \Theta)}\right\vert^2 - 1\right)^2, 
\end{multline}
and an optimal configuration of $\Theta$ can be obtained by minimizing $\chi^2(\Theta, \mathbb D_{\rm train})$ using classical standard optimizers~\cite{hansen_cma_2006, byrd_lbfgsb_1995}.

The performance of supervised-learning models is not measured by its capability to learn the training dataset, but its generalization power. Then, the quantity of interest is the accuracy $\mathcal A$. This quantity counts the number of correct guesses over samples on an unlearned test dataset $\mathbb D_{\rm test}$. The guessed class is
\begin{equation}
    c_g(\vec x, \Theta) = \arg\max_{\bar{c}} \left\vert \braket{\phi_c}{\psi(\vec x, \Theta)}\right\vert^2
\end{equation}
and the accuracy $\mathcal A(\Theta; \mathbb D_{\rm test})$ is  
\begin{multline}\label{eq:accuracy}
    \mathcal A(\Theta; \mathbb D_{\rm test}) =\\ \frac{1}{N_{\rm test}}\sum_{(\vec x, c)\in \mathbb D_{\rm test}} {\rm bool}(c_g(\vec x, \Theta)= c) , 
\end{multline}
with ${\rm bool}(a = b)$ is $1$ if $a = b$, and $0$ otherwise. The value of $\mathcal A(\Theta; \mathbb D_{\rm test})$ is bounded between $0$ and $1$, being $1$ the optimal result.

\subsection{Optimization}~\label{sec:optimization}

\begin{figure*}
    \centering
    \subfigure[Circuit compression\label{fig:reduce_depth}]{
    \resizebox{.65\linewidth}{!}{
    \hspace{.0cm}~\Qcircuit @C=.8em @R=.7em{
    \push{\rm Theory~\qquad} & \lstick{\ket 0} & \qw & \gate{R_y(\theta^y_1)} & \qw & \gate{R_z(\theta^z_1)} & \gate{R_y(\theta^y_2)} & \qw & \gate{R_z(\theta^z_2)} & \gate{R_y(\theta^y_3)} & \qw & \gate{R_z(\theta^z_3)} & \qw & \qw & \qw & \push{\cdots}\\
    & & & & \\ 
    & & & & & \\
    & & & & \\
    \push{\rm Experiment~\qquad} & \lstick{\ket 0} & \qw & \gate{\mathcal R(\pi / 2, \theta^y_1)} & \qw & \qw & \gate{\mathcal R(\bar\theta^z_1,\theta^y_2)} & \qw & \qw & \gate{\mathcal R(\bar\theta^z_2,\theta^y_3)} & \qw & \qw & \qw & \qw & \qw & \push{\cdots}
    \protect\gategroup{1}{4}{1}{6}{1em}{--}
    \protect\gategroup{1}{7}{1}{9}{1em}{--}
    \protect\gategroup{1}{10}{1}{12}{1em}{--}
    \protect\gategroup{1}{4}{5}{4}{.7em}{(}
    \protect\gategroup{1}{4}{5}{4}{.7em}{)}
    \protect\gategroup{1}{6}{5}{7}{.7em}{(}
    \protect\gategroup{1}{6}{5}{7}{.7em}{)}
    \protect\gategroup{1}{9}{5}{10}{.7em}{(}
    \protect\gategroup{1}{9}{5}{10}{.7em}{)}
    \protect\gategroup{1}{12}{5}{12}{.7em}{(}
    }
    }}
    ~\hfill
    \subfigure[Measurement\label{fig:measurement}]{
    \resizebox{.25\linewidth}{!}{
    \hspace{.0cm}~\Qcircuit @C=.5em @R=.7em{
    \push{\hspace{1mm}} & \lstick{\ket 0} & \qw & \gate{U(\vec x, \Theta)} & \qw  & \gate{V^\dagger_c} & \qw & \meter & \rstick{\Rightarrow P_0} \\
    & \\& \\& \\ &
    }}\hspace{1cm}}
    \begin{textblock}{4.5}(1.9, -.8)
    \small$
    \bar\theta_1^z = \theta_1^z + \pi / 2$
    \end{textblock}
    \begin{textblock}{4.5}(3.8, -.8)
    \small$
    \bar\theta_2^z = \bar\theta_1^z + \theta_2^z$
    \end{textblock}
    \begin{textblock}{4.5}(5.75, -.8)
    \small$
    \bar\theta_{n}^z = \bar\theta_{n-1}^z + \theta_n^z$
    \end{textblock}
    \begin{textblock}{4.5}(9, -.8)
    $
         P_0 = \vert \bra 0 V^\dagger_c U(\vec x, \Theta) \ket 0 \vert^2 = $ \\
     \hspace{15mm}$=\left\vert \braket{\phi_c}{\psi(\vec x, \Theta)}\right\vert^2$
    \end{textblock}
    \caption{Methods applied in the classifier to optimize the implementation. a) Depth reduction of layers. Layers are grouped in dashed lines. The rotations $R_y$ from a layer and $R_z$ for the previous one are combined to be executed by means of only one $\mathcal R(\cdot, \cdot)$ operation, thus effectively reducing the depth of the circuit. This is indicated by brackets. For consistency, a redefinition of parameters in the $R_z$ rotations must be carried, detailed in the figure. b) Measurement of the fidelity between two states. The state $\ket{\psi(\vec x, \Theta)} = U(\vec x, \Theta)\ket 0$ is compared against $\ket{\phi_c} = V_c\ket 0$. The probability of measuring the $\ket 0$ state at the end of this circuit equals the relative fidelity.}
    \label{fig:tricks}
\end{figure*}
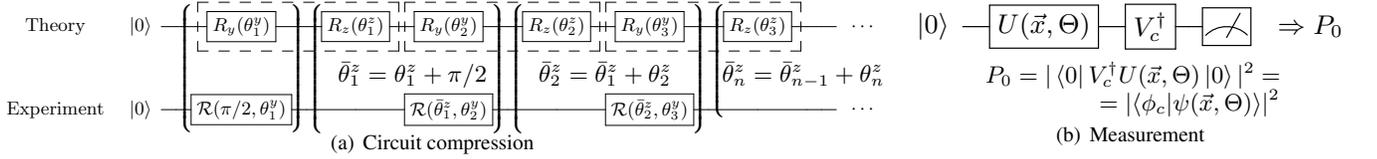

The classifier proposed in this work is trained in two steps. First, given a $\mathbb D_{\rm train}$ dataset, a simulated version of the model is created and trained using the $\chi^2$. The optimization returns a set of parameters $\Theta^{\rm sim}$. Using exact simulations permits to circumvent difficulties in the training process due to noise and uncertainties. 

In a second step, the obtained parameters are ported to the experimental classifier. A relevant improvement can be obtained by fine-tuning the given parameters when executed on the quantum hardware. In most scenarios, parameters for simulation and experiment do not exactly match each other. The difference may stem from the experimental inaccuracy when implementing optimal angles and occasional loss of information due to collisions with background molecules. This second optimization step polishes the experiment to reduce any systematical error occurring during the execution of the quantum circuit. The parameter set $\Theta^{\rm sim}$ is taken as starting point to explore the parameter space in its vicinity. A new set of parameters reducing experimental inaccuracies is obtained as $\Theta^{q} = \Theta^{\rm sim} + \delta\Theta$. The figure of merit to be optimized in this case is $\mathcal{A}^{q}$. This second step is currently available for quantum devices only if the loss function in the parameter space near the vicinity of the minimum is shallow and large deviations from the optimal parameters translate into small changes in the loss functions. See Fig.~\ref{fig:fig1} for a scheme describing the two-step optimization performed.

The full scan of the space of parameters is very time expensive in the present experiments. In this work, the optimization of parameters on the experiment is performed sequentially. First, the two first parameters are taken, while all others are fixed. A two-dimensional scan is performed for those parameters in the nearby around $\Theta^{\rm sim}$. The pair of parameters is then modified for improving the obtained accuracy. The next step is performed on the second and third parameter. With each iteration, one new parameter is taken into account. This method returns substantial improvements only on the first layers and can be stopped after few iterations without a significant loss of performance. 

Notice that the two-step optimization procedure returns three different values of the accuracy $\mathcal A$ to be measured, namely 
\begin{itemize}
    \item[{\it i)}] $\mathcal{A}^{*} $ obtained by running an exact simulation of the quantum circuit with the optimal parameters $\Theta^{\rm sim}$
    \item[{\it ii)}] $\mathcal{A}^{\rm sim} $ run on the QPU with the same parameters $\Theta^{\rm sim}$
    \item[{\it iii)}] $\mathcal{A}^{q} $ obtained with the parameters $\Theta^{q}$ after the second optimization step.
\end{itemize}   

In addition, for every classification problem, the obtained accuracies can be compared with genuinely classical models. It is expected that experimental errors will deteriorate the ideal performance of the computer. However results remain very competitive in the experimental setup.

Both the two-step optimization and the accuracies related to classification problems on experiments are features surpassing previous results in Ref.~\cite{perez-salinas_one_2021}. In previous works, parameters are directly ported from simulation setups to the experiment. A reasonable degradation in the final results is observed, but there is no further study on how to improve the hardware implementation of the algorithm. With regard to the accuracies, the previous work does not need to generalize from training to test dataset, but only show the flexibility to mimic a given behavior. In the classifier exposed in this work, generalization is a requirement successfully achieved. 

\subsection{Optimal hardware control}\label{sec:opt_hardware}

In order to improve the performance of the classifier to the limit of the capabilities of the present QPU, some features were implemented specifically for this experiment.

First, the error in the applied gates is reduced by cleverly choosing the application of axis. In the ion qubit, active rotations $R_x$ and $R_y$ are realized by the interaction of the resonant laser with the qubit at the specified frequency during a time proportional to the rotation angle. On the other hand, $R_z$ can be implemented by a combination of the other two active gates or by varying the qubit energy~\cite{PhysRevA.96.022330, Maslov_2017}. Both these methods are error prone as the qubit-light interaction is switched on for certain time. We chose instead to perform the $R_z$ by changing the laser phase without interacting with the ion~\cite{PhysRevA.77.012307}, thus the resultant error is limited by only the $R_y$ gate in each layer. 

Second, the depth of the circuit can be effectively halved by combining $R_z(\gamma)$ and $R_y(\delta)$ gates into a single gate $\mathcal R(\gamma, \delta)$ which applies a rotation of angle $\delta$ around the axis $(\gamma, 0)$, defined in spherical coordinates. 

This rotation around and arbitrary axis is defined as
\begin{equation}\label{eq:matrixR}
    \mathcal R(\gamma, \delta) = \begin{pmatrix}
            \cos\frac{\delta}{2}   & -ie^{-i\gamma}\sin\frac{\delta}{2} \\
            -ie^{i\gamma}\sin\frac{\delta}{2}  & \cos\frac{\delta}{2}       
    \end{pmatrix},
\end{equation}
which can be understood as 
\begin{equation}
\mathcal R(\gamma,\delta) = R_z(\gamma) R_x(\delta) R_z(-\gamma).
\end{equation}
Thus, it is possible to relate $R_x(\delta) = \mathcal R(\gamma, \delta)$ and $R_y(\delta) = \mathcal R(\pi / 2,\delta)$. Taking the Ansätz from Sec.~\ref{sec:reuploading} it is possible to decompose those gates into $\mathcal R(\gamma,\delta)$ operations to effectively reduce the number of operations. The first $R_y(\cdot)$ gate must be applied on its own. For the $n$-th step, the $R_z$ rotation from the $(n-1)$-th layer and the $R_y$ from the $n$-th layers can be composed into only one gate. As a consequence, the $n$-th $\mathcal{R}$ gate must adjust its parameters to 
\begin{eqnarray}
    \bar\theta_{1}^z &=& \pi / 2 + \theta_1^z \\
    \bar\theta_{n}^z &=& \bar\theta_{n-1}^z + \theta_n^z.
\end{eqnarray}See Fig.~\ref{fig:reduce_depth} for a schematic description of this process.

This reduction can only be applied to pairs of gates and not to the entire circuit. The reason is that the data $(x)$ and tunable parameters $(\vec \theta)$ from Eq.~\eqref{eq:gen_circuit}come into the circuit through linear operations. Thus, two gates can be easily combined by classically performing this operation. On the other hand, the chain of repeated gates gives rise to complex non-linear behaviors that cannot be easily calculated by classical means. 

The measurement process to construct the cost function involves the quantities $\vert \braket{\phi_c}{\psi(\vec x, \Theta)}\vert$ as described in Eq.~\eqref{eq:label_states}. A direct method to obtain this quantity is by performing tomography and obtaining an indirect computation of the fidelity. However, tomography demands a large amount of measurements and resources to obtain an accurate result. Instead, the fidelity is computed by direct comparison of the result state and the label states. For this purpose, the label states are defined by means of a simple gate $V_c$ as
\begin{equation}
    \ket{\phi_c} = V_c\ket{0} = R_z(\eta) R_y(\lambda) \ket 0,
\end{equation}
where the parameters $\eta, \lambda$ are problem-dependent. They must be tuned to accomodate several states as orthogonally as possible accross the space in the Bloch sphere. This is a problem on its own, but in some cases, including those tackled in the present work, the solution is trivial \cite{perez-salinas_data_2020}. Thus, fidelity can be measured by comparing the output and label states as 
\begin{multline}
    \vert \braket{\phi_c}{\psi(\vec x, \Theta)}^2 = \vert \bra{0} V^\dagger_c U(\vec x, \Theta) \ket 0\vert ^2 = \\ = P_0(V^\dagger_c U(\vec x, \Theta))
\end{multline}
where the last term corresponds to the probability of measuring the state $\ket 0$ after executing the circuit composed by both operators. The corresponding scheme can be viewed in Fig.~\ref{fig:measurement}.

\section{Results}\label{sec:results}

\begin{figure*}[t!]
     \begin{center}
   \includegraphics[width=.85\textwidth]{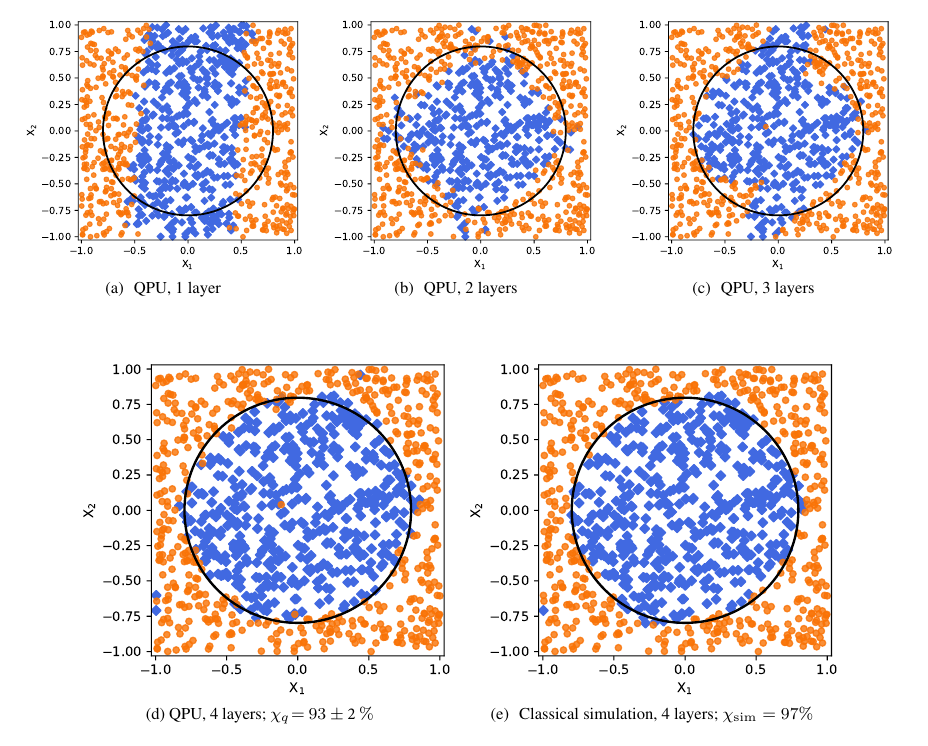}
        
    \end{center}
\caption{Results for the data set of the ion-trap re-uploading classifier as compared to classical simulation. $1000$ random data points are tested, and the guessed classes are depicted in blue diamonds and orange circles respectively for points inside and outside the circular boundary, in solid line. In (a-d), the depth of the circuit is increased from 1 to 4 layers, showing gradual improvements of the classification. The result of the $4$-layer QPU classifier (d) is compared with the equivalent $4$-layer simulation (e) for benchmark. Notice that the border between classes in the experimental results is not as sharply defined as in the simulated classification due to the uncertainty of the quantum measurements and systematic errors.
}
   \label{fig:fig2}
      
\end{figure*}

The problems solved in this work consist in separating points in a feature space depending on its location. The classes are defined by different curves acting as boundaries. The problems proposed can only be solved if the classifier is flexible enough as to map close points to distant areas of the Hilbert space. Experimental results are presented with an increasing level of difficulty. For every instance, we compare the accuracy of the simulated and the experimental classifiers $\mathcal{A}^{*}$ and $\mathcal{A}^{\rm sim}$ respectively, and benchmark them against classical algorithms. 
In all cases, a training dataset of 200 randomly distributed points is used for optimization. The accuracy is tested against 1000 different points. This way, generalization is also tested on unseen data.
In two cases, due to its high computational cost, the experimental optimization step to obtain $\mathcal{A}^{q}$ is carried. A full analysis is deferred to a separate publication. Images illustrating results are included for some examples. Additional images can be found in the Appendix. 

\bigskip 

\paragraph{Two dimensional binary classification.} 

A binary classifier is designed to learn the difference between points in a plane inside $(0)$ and outside $(1)$ a circle. The circle is centered at the middle of the plane and has half the total area, so that the instances belonging to the two classes are balanced. A random classifier would only guess 50$\%$ of test data properly. 

Fig.~\ref{fig:fig2} shows experimental results for this classification problem for an increasing number of layers, and a comparison between the test data as classified by the QPU and an ideal classical qubit simulator. Here, each data point is depicted by its coordinates $(x_1, x_2)$ and coloured according to its binary classification, namely blue for inside and orange for outside the circle, shown as a black solid line.
The figure shows the gradual improvement of classification as more layers are added, confirming results from Ref.~\cite{perez-salinas_data_2020}. For 4 layers, the classification accuracy for the QPU is $\mathcal{A}^{\rm sim} = 93\pm{2}\%$, slightly lower than its classically simulated counterpart $\mathcal{A^{*}}= 97\%$. The error in $\mathcal{A}^{\rm sim}$ refers to the standard deviation of $10$ repeated trials performed on the same dataset. Underlying systematic uncertainty leads to differents values of the accuracy.

It is interesting to note a difference in the behavior of both simulated and quantum classifiers. In the simulation, Fig.~\ref{fig:fig2}(e), the guessed boundary between classes is sharply defined, even though the guesses do not match exactly the data and the circle is slightly deformed. The results on the experimental data show uncertainty in the determination of data. See for instance the higher part of the circle. There are several points guessed to belong to different classes interspersed in a small area. The origin of this phenomenon is the inherent sampling uncertainty of the quantum device.

\begin{figure*}[ht!]
    \begin{center}
            \includegraphics[width=.95\linewidth]{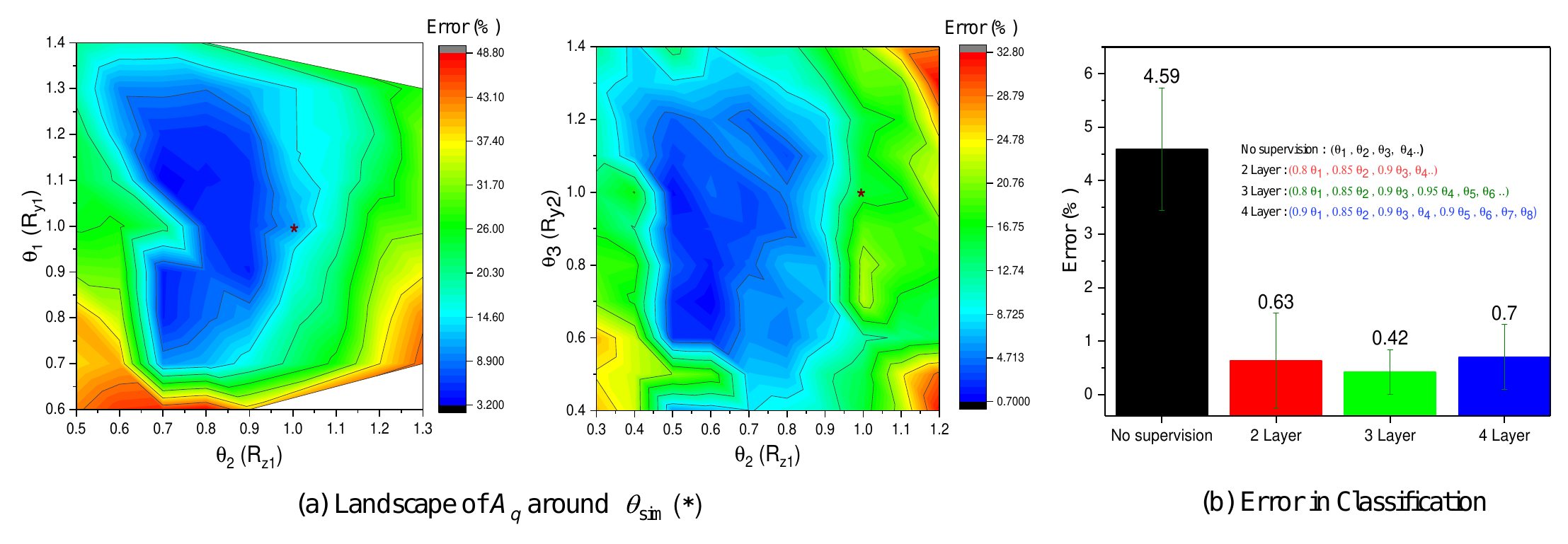}
\end{center}
\caption{Results for the experimental second step for optimization in the circle problem. The classifier is first trained on a classical simulator, obtaining a set of parameters $\Theta^{\rm sim}$. In a second step, the vicinity of this point in the parameter space is explored using the QPU. A new set of parameters $\Theta^q = \Theta^{\rm sim} + \delta\Theta$ is selected to optimize the accuracy of the classifier by mitigating experimental errors. (a) Error surface for the deviation of the optimal parameters from the $\Theta^{\rm sim}$~$(\star)$ for selected parameters. These plots correspond to the 2-layer classifier from Fig.~\ref{fig:fig2}. (b) Betterment of the accuracy in the 3 and 4 layer QPU classifier, as compared to simulated results. The error in classification is about $5\%$ when $\Theta^{\rm sim}$ is used. On the contrary, if $\Theta^q$ are used, the errors are reduced below $1\%$. The improvement in the first two layers are the most prominent.}%
\label{fig:supervised}

\vspace{2mm}

     \begin{center}
       
           \includegraphics[width=.8\textwidth]{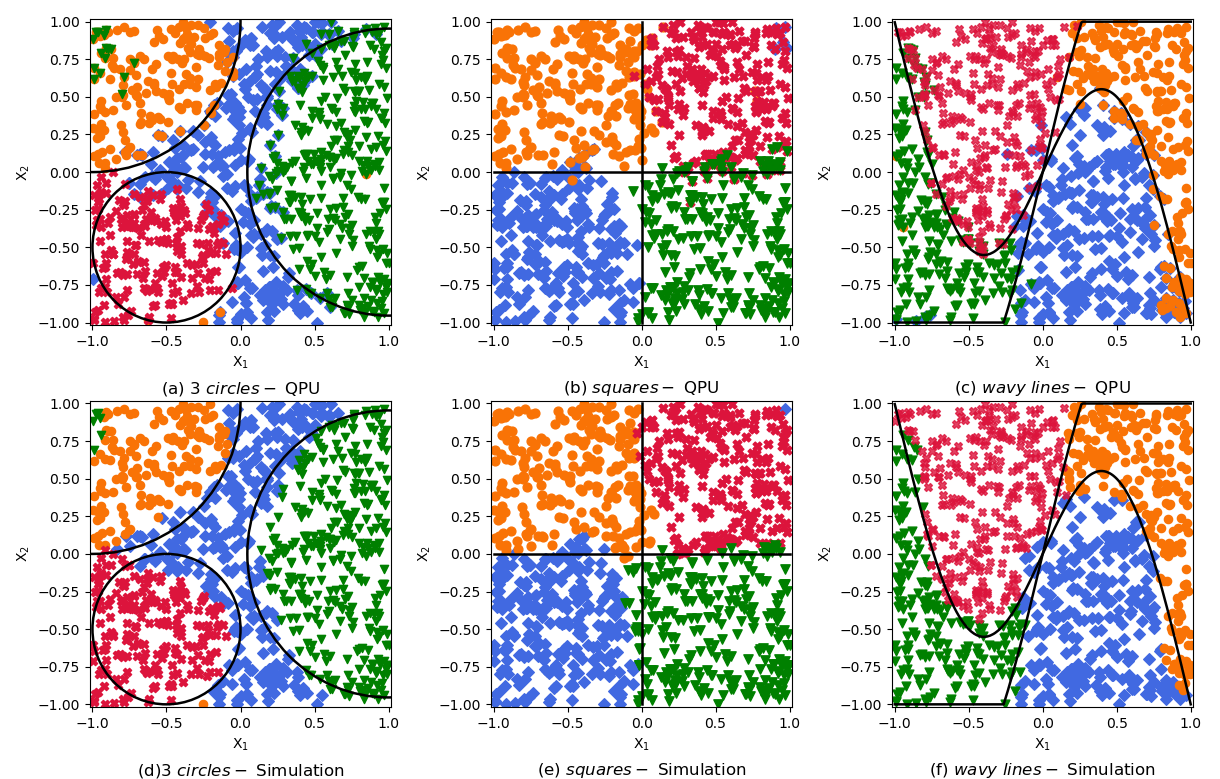}
       
    \end{center}
 \caption{Multi-class classification. The columns show, from left to right, {\sl 3 circles}, {\sl squares} and {\sl wavy lines} problems. The top row corresponds to experimental QPU results, while the bottom row corresponds to classical simulations. Results include $1000$ random data points, the classes are depicted in different colors and symbols, and the boundaries are defined by solid black lines. Notice that the border between classes in the experimental results is not sharply defined, unlike in the simulated classification. This difference is due to the uncertainty of the quantum measurements.}%
   \label{fig:multiclass}
\end{figure*}

In this example, the experimental second optimization step is also performed. Figs.~\ref{fig:supervised}(a) show the landscape of the accuracy for a specific subset of parameters in the vicinity of the optimum point as provided by the classical simulation $\Theta^{\rm sim}$. In this case, only three parameters effectively contribute to the final result, and thus the cost of searching optimal experimental configurations using a scanning technique is manageable. 

The error in classification of the experimental implementation with respect to the exact simulation is also depicted in Fig.~\ref{fig:supervised}(b). Upon correction, the deviation between $\mathcal{A}^q$ and $\mathcal{A}^{*}$ nearly vanishes within the experimental uncertainty. After optimization, we obtain an improvement by more than $5\%$ that brings accuracy of QPU to be as good as the simulator. This is better understood, by looking at the optimized error in Fig.~\ref{fig:supervised}(b) with respect to the simulation plotted as a function of the number of layers in the supervised learning process. In a realistic situation, the loss of information due to collision (small percentage to the systematic error) can also be corrected by the observation of the emitted light level at the cost of longer operation time. Experimental errors are estimated separately. A detailed discussion on experimental errors is carried in App.~\ref{sec:errors}.

Additionally, two more $2$-dimensional binary classification problems, namely  {\sl non-convex} and {\sl crown}, can be seen in Fig.~\ref{fig:nonconvex_crown} of App.~\ref{sec:additional} for 4 layers. In this case, the final accuracies ($\mathcal{A}^{\rm sim}$) / ($\mathcal{A}^{*}$) are {\sl non-convex}:($92\%$) / ($95\%$); {\sl crown}: ($87\%$) / ($92\%$).

\bigskip
\paragraph*{Higher dimensional classification.} A single qubit is capable to address classification problems in an arbitrary number of dimensions \cite{perez-salinas_data_2020,perez-salinas_one_2021,schuld_effect_2021}.
In order to test a classification performance for high-dimensional data, we extend the 2D-{\sl circle} problem to 3D-{\sl sphere} and 4D-{\sl hypersphere} problems. The statement of the problem is equivalent and only the radius of the boundary is modified to accommodate the requirement that every class corresponds to half the feature space. Results for the {\sl hypersphere} classification are depicted in Fig.~\ref{fig:hyper_4d} of App.~\ref{sec:additional}. The error in the experimental setup using $\Theta^{\rm sim}$ can be further reduced from $\sim 13\% $ to $\sim 2\% $ after the experimental optimization step is performed.

\bigskip

\paragraph*{Multi-class classification.} The single qubit QPU can also maximally separate three, four or more different classes \cite{perez-salinas_data_2020}. To prove it experimentally, we provide an example of a three-class classification, {\sl tricrown}, and three different examples of a four-classes problem, namely, {\sl 3 circles}, {\sl squares} and {\sl waves}, in a 2D feature space. The results for all examples are depicted in Fig.~\ref{fig:multiclass} and in Fig.~\ref{fig:tricrown} of App.~\ref{sec:additional} for $4$ layers for QPU and classical simulator. Results are respectively: {\sl tricrown}: ($95\%$) / ($91\%$); {\sl 3 circles}:($85\%$) / ($90\%$); {\sl squares}:($93\%$) / ($97\%$); {\sl waves}:($90\%$) / ($94\%$). 

These datasets present different hardness for classification~\cite{Class_complexity}. The {\sl $3$ circles} feature is composed of $3$ different and separated classes, that is the circles, and a fourth class filling the space in between. Thus, the classifier is forced to separate the space corresponding to different classes. In the {\sl squares} case, all classes are equivalent and connected to each other. For {\sl waves}, the difficulty of non-convex datasets is added. Despite various level of difficulties in classification, the experimental classifier succeed in solving all with nearly the same percentage.

\paragraph*{Summary of results.} Table~\ref{tab:table1} presents a summary of the accuracies obtained by the experimental quantum classifier as compared to the simulated results serving as the starting point for the experimental optimization. Each problem was solved using an independent Ansatz (A, B) specified in the table. We also benchmark the quantum results against classical models whose complexity is comparable. To be precise, we have considered a single-hidden-layer neural network and a support vector machine classifier. In the case of Neural Networks, several activation functions were tested for every problem, and the result depicted in Tab.~\ref{tab:table1} is the best performing one. The number of neurons in the hidden layer is chosen to match the number of parameters in the quantum classifier. In the case of support vector machine classifier, different kernels are used, and again the best result is retrieved. Notice that, in this case, it is not possible to tune the number of parameters available to carry the classification since they depend on the size of the training set. This model is however shown for completeness. In all classical cases, the computations were done using the {\tt scikit-learn} python package~\cite{scikit-learn} and following pre-defined models.

\begin{table}[t!]
\centering
\begin{tabular}{|c||c|c||c|c|c||c|}\hline
\multirow{2}{*}{Problem (\# classes)} & \multicolumn{2}{c||}{\bf Classical} & \multicolumn{3}{c||}{\bf Re-uploading} & \multirow{2}{*}{Ansatz} \\
 & $\mathcal{A}^{NN}$ & $\mathcal{A}^{SVM}$ & $\mathcal{A}^{*}$ & $\mathcal{A}^{\rm sim}$ & $\mathcal{A}^{q}$ &\\ \hline\hline
 Circle \hfill (2) & 98 & 96  & 97 & 93 & 96& A\\
 Crown \hfill (2) & 71 & 82 & 92 & 87 && B \\  
 Non-Convex \hfill (2)& 98 & 79 & 95 & 92 && B\\
 Sphere \hfill (2) & 95 & 91 &74 & 66 && A\\
 Hypersphere \hfill (2) & 76 & 92 & 75 & 64 & 73& A\\ 
 Tricrown  \hfill (3) & 97 & 83 & 95 & 91 && A\\ 
 3 circles \hfill (4) & 93 & 92  & 90 & 85 && B\\
 Squares \hfill (4) & 99 & 95 & 97 & 93 && A\\ 
 Wavy Lines \hfill (4)& 99 & 89 & 94 & 90 && A\\ \hline

\end{tabular}
\caption{Comparison between the accuracies, in $\%$, of the single-qubit re-uploading quantum classifier and two well-known classical classification techniques, namely single-hidden-layer neural networks (NN) and support vector machines (SVM). The experimental data and its simulated analogue is provided here with 4 Layer and 100 repetitions on the quantum part, and its equivalent in number of parameters for the neural network. The uncertainty of experimental data is $\pm 2\%$. The error refers to the standard deviation of $10$ repeated trials performed on the same dataset and it implies that underlying systematic uncertainty leads to an uncertainty of the accuracy, see App.~\ref{sec:errors} for a detailed analysis. Only two cases have been further optimized using an exploration done only with the quantum device.}
\label{tab:table1}
\end{table}

In light of these results it is possible to see that the performance of the quantum classifier is comparable to that of classical classifiers when the training dataset has a low feature dimension. On the contrary, when datasets with more features are considered, the performance degrades but are still comparable to simulated results. This limitation can be addressed by increasing the flexibility of the model, as it was shown in Ref.~\cite{perez-salinas_data_2020}. Two ways to achieve higher flexibility is to either add more layers to the quantum classifier or to change the encoding scheme to one with more degrees of freedom. 

\section{Discussion} \label{sec:discussion}

We have experimentally implemented a single-qubit quantum supervised classifier on a QPU. The QPU is based on an ion trap platform, which is known for high fidelity gates~\cite{Yum:17, PhysRevA.77.012307, PhysRevLett.117.060505}. This high fidelity allows to implement a quantum classifier based on the recently proposed re-uploading algorithm~\cite{perez-salinas_data_2020, perez-salinas_one_2021}. To the best of our knowledge, the experiment here reported is the first implementation of problems of this kind in the field of classification with minimal quantum resources, surpassing the experimental results from the previous work~\cite{perez-salinas_one_2021}. This work constitutes the first experimental proof that the re-uploading scheme may become part of QML strategies for classification.

To enhance the performance of the algorithm, the classifier is trained in two steps, first using a classical simulator, and then at the gate level of the quantum device. This second step permits to mitigate possible systematic errors from the quantum device. With this method, the accuracy of the classifier on the quantum experiment is enhanced up to $5-10\%$ depending on the problem. The resultant high performance of the QPU allowed comparable outcome to that of a classical simulator and classical classifiers of equivalent complexity in the number of parameters used.

The examples of the datasets provided here include rudimentary but non-trivial classification tasks that were successfully solved, including binary, multi-class and high dimensional datasets. To complete the bench-marking, we showed that our QPU can not only classify multi-class and higher dimensional feature maps but also shows competing results in classifying non-convex and linear feature maps. Both experimental and simulated results are benchmarked against well-established classical methods. The performances for quantum and classical cases are comparable, see Tab.~\ref{tab:table1}. 
These results are aligned with the analysis from Ref.~\cite{perez-salinas_one_2021} where it is demonstrated that single-qubit re-uploading circuits with $N$ layers are formally equivalent to single-hidden-layer neural networks with $N$ neurons in the hidden layer.

While other approaches explore the possibilities to classify complex datasets by implementing a variety of quantum models, 
the aim of the re-uploading scheme is to compare their performance with classical procedures of similar complexities. This work demonstrates experimentally that a minimalistic quantum system, namely a single qubit, is capable to perform non-trivial classification tasks. This may bring advantage on quantum algorithms used as a subroutine to perform QML tasks while maintaining the requirements of quantum resources to its minimum. 

\acknowledgments
This research is supported by the Quantum Engineering Program under National Research Foundation (Award No. QEP-P6) and CQT Core grant. A.P.-S. acknowledges financial support from Secretaria d'Universitats i Recerca del Departament d'Empresa i Coneixement de la Generalitat de Catalunya, co-funded by the European Union Regional Development Fund within the ERDF Operational Program of Catalunya (project QuantumCat, ref. 001-P-001644).

\bibliography{bibliography}

\appendix

\section{Error analysis}\label{sec:errors}

\begin{figure*}[t!]
    \begin{center}
            \includegraphics[width=.9\linewidth]{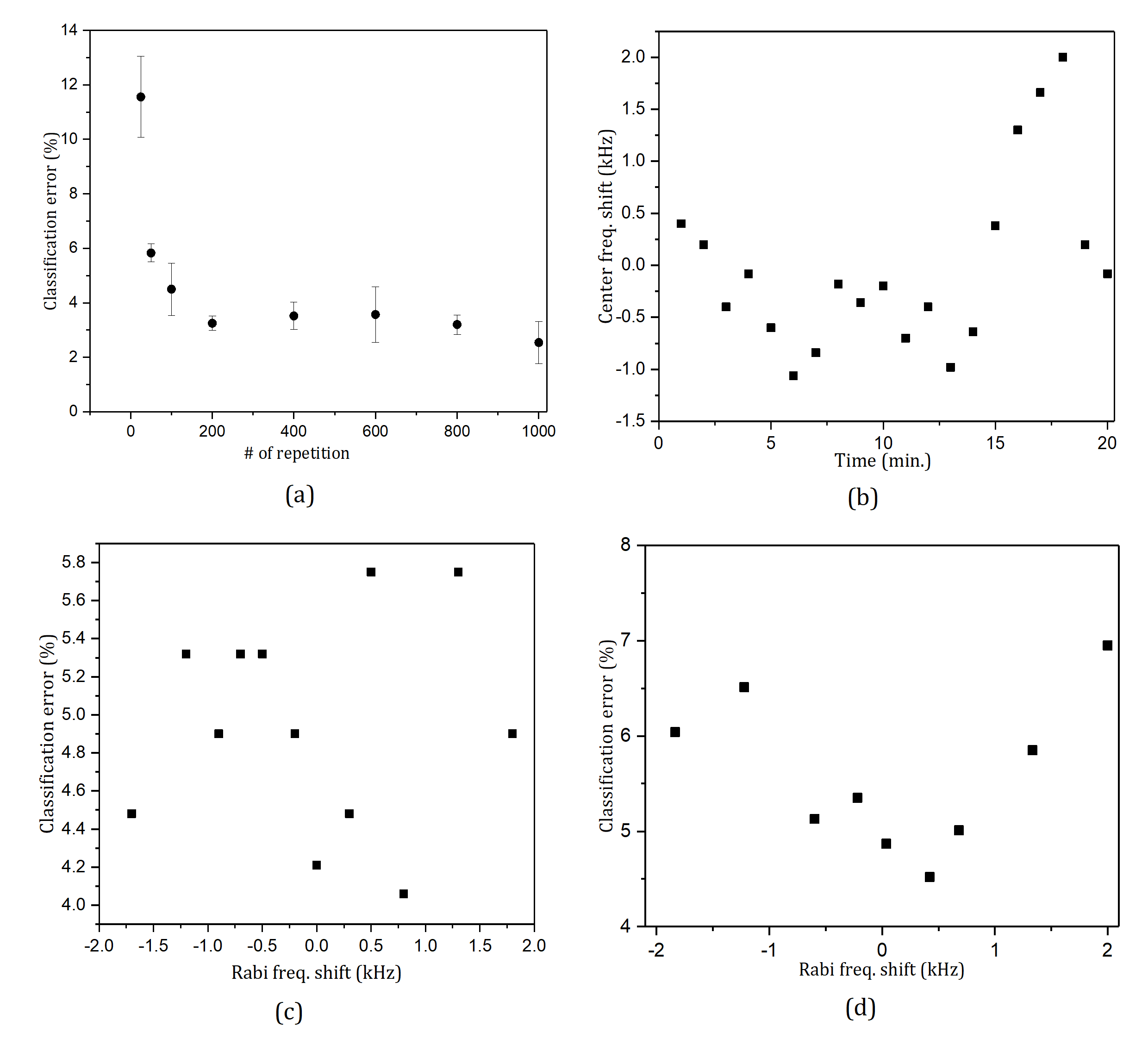}
    \end{center}
\caption{Systematic error analysis: All the results shown here are related to the binary classification of circle as in Fig.~(\ref{fig:fig2}). The errors are classification error. (a) The classifier error as a function of the number of repeated experiments. The error bar at each point corresponds to $1$ standard deviation of a number of repeat measurements for same number of repeated experiments under the same condition. The exact number of repeat measurements varies between $5$ and $10$. (b) Variation of the resonance frequency as a function of time. The range of Rabi frequency fluctuation within a typical experimental time of $\le 10$ min. is about $2$ kHz. (c) Error in binary classification of a circle feature with the variation of laser frequency detuning measured in terms of the Rabi frequency. The variation in the value of classification error is  about $2\%$ within the experimental time of $\sim 10$ min. (d) The same plot as in (c) but by varying the laser power measured in terms of the Rabi frequency.}%
  \label{fig:error}
\end{figure*}

The accuracy of the data re-uploading algorithm primarily relies on the the fidelity of individual single qubit rotation gate. The gate, as explained in the main text, is operated by controlling the laser phase and interaction time, while keeping the intensity at the ion position and the frequency of the laser constant. Therefore, the residual error in the gate operation is reflected in the accuracy of the classifier as define in eq.~(\ref{eq:accuracy}). In Eq.~\eqref{eq:matrixR}, the rotation angles $\gamma$ and $\delta$ are related to physical quantities as
\begin{eqnarray}
\gamma & = & (\Delta/\hbar) t_{op}+\delta\phi\\ 
\delta & = & (\Omega/\hbar) t_{op},
\end{eqnarray} where $\Delta$ is the laser detuning, $t_{op}$ is the operation time of gates, $\Omega'$ is the modified Rabi frequency and $\delta\phi$ is the relative phase of the laser with respect to the qubit. The modified Rabi frequency $\Omega' = \sqrt{\Omega_0^2+\Delta^2}$ with $\Omega_0$ denoting the resonant Rabi frequency. Furthermore, the resonant Rabi frequency is proportional to the square root of the intensity, $I_0$, at the ion position. Therefore each of the independent variables $\delta\phi$, $t_{op}$, $\Delta$ and $I_0$ contributes to the error in a rotation gate, thus influencing the accuracy of the quantum data re-upload classifier. We have characterized these factor separately using the {\it circle} classification problem as a test-bed. The errors shown in Fig.~(\ref{fig:error}) are obtained as in-accuracy percentage of classification as compared to the simulation results. In the following, the detail of the influence of each of these parameters on the accuracy of the classifier is discussed:

\begin{itemize}
    \item[1.]{\bf Phase:} The DDS controls the RF phase of the AOM which determines the relative phase of the laser. Each DDS is synchronized to a rubidium atomic clock which is accurate to one part in $10^{10}$ and thus contributes negligibly to the phase error. The DDS is however triggered by the FPGA which has time jitter below $10~$ns leading to phase noise on the qubit below $0.1\%$ for a Rabi $\pi$ time of $12~\mu$s.

    \item[2.]{\bf Interaction time:} The laser-qubit interaction time is determined by the FPGA which is precise to $1~$ns. Therefore the contribution to the accuracy of the classifier is less than $0.01\%$. However, due to the time jitter below $10~$ns, its contribution to the accuracy is below $0.1\%$. Occasional collision with the residual background gas molecule during the interaction time leads to a projection to the state $\ket{0}$, thus losing the final state information and hence error in the classification. Usually this error becomes smaller with larger statistics. 
    
    \item[3.]{\bf Laser-qubit detuning:} The detuning of the laser with respect to the qubit frequency denoted by $\Delta$ modifies the Rabi frequency. In order to quantify the influence of any unwanted fluctuations of the detuning on the classifier accuracy, we first quantified the range of Rabi frequency fluctuation within the experimental time of about $10~$min. However, to ensure that our classifier accuracy is limited by systematic errors not by statistical, we measured the statistical error in the classification problem by repeating the experiment for each data point between $25$ to $1000$ times (see Fig.~(\ref{fig:error}a)). The error (or in-accuracy) decreases from $12\%$ to about $4\%$ for $100$ repetitions and then stay nearly the same, limited by systematic. The fluctuation of the laser frequency with respect to the atomic resonance is captured over a time period of $20~$min (twice the duration of an experiment) as plotted in Fig.~(\ref{fig:error}b). The random variation of the Rabi frequency over time is mostly caused by the magnetic field noise as we have separately measured the laser frequency drift to be $\le 5~$kHz/$24$ hrs~\cite{Yum:17}. To minimize the impact of the residual magnetic field noise, we use electronic levels ($\Delta m = 0$) that are weakly sensitive to such noise. In addition, the detuning also indirectly influence the modified Rabi frequency. To check its influence, we varied the Rabi frequency, shown in Fig.~\ref{fig:error}(c), by varying the detuning within a $2~$kHz range (as expected from the Fig.~(\ref{fig:error}b)). The result shows below $5\%$ accuracy for the classifier when operating for $10~$min.
    
    \item[4.]{\bf Laser intensity:} The Rabi frequency is fixed by setting the power and frequency of the laser at the start of the experiment. Any change in the Rabi frequency during the experiment, therefore leads to error in applied qubit rotation angle. Therefore, the accuracy of the classifier, depends on the Rabi frequency fluctuations due to intensity fluctuation apart from detuning as discussed earlier. The intensity is influenced by two factors namely: (a) laser power noise and (b)  laser beam pointing error. In our experiment, the laser beam is tightly focused on the ion by a high Numerical Aperture (NA$\sim$0.4) in-vacuum lens. To obtain high intensity at the ion position, the light is focused tightly to about $10\mu$m beam waist. In order to capture the influence of laser power variations on the classification error, we varied power (plotted in terms of modified Rabi frequency) in Fig.~(\ref{fig:error}d). Thus it is seen that the influence of intensity noise accounts to $5\%$ error in accuracy. Thus, to avoid the influence of Rabi frequency fluctuation within the experimental time of $\sim 10~$min, we reduce the Rabi frequency from $312~$kHz to $40~$kHz such that the absolute error also reduces. This leads to an overall error of only $2\%$ on the classifier output.       
\end{itemize}

\section{Additional results}\label{sec:additional}

 \begin{figure*}[h!]

    \centering
    \subfigure[\hspace{1mm} {\sl Non-convex} - QPU\label{fig:fig3a}]{
    \includegraphics[width=.23\linewidth]{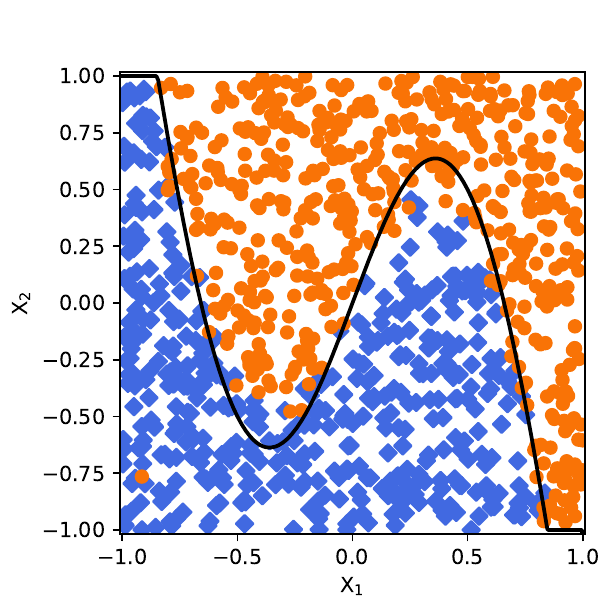}}
    \subfigure[\hspace{1mm} {\sl Non-convex} - Simulation\label{fig:fig3c}]{
    \includegraphics[width=.23\linewidth]{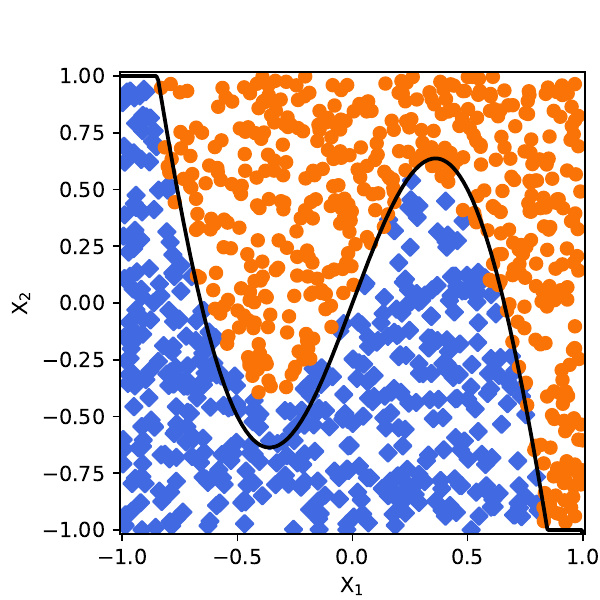}
    }
    \subfigure[\hspace{1mm} {\sl Crown} - QPU\label{fig:fig3b}]{
    \includegraphics[width=.23\linewidth]{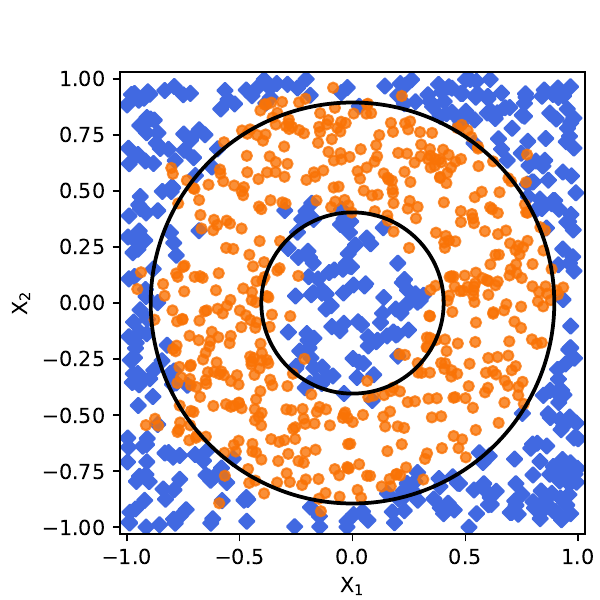}
    }
    \subfigure[\hspace{1mm} {\sl Crown} - Simulation\label{fig:fig3d}]{
    \includegraphics[width=.23\linewidth]{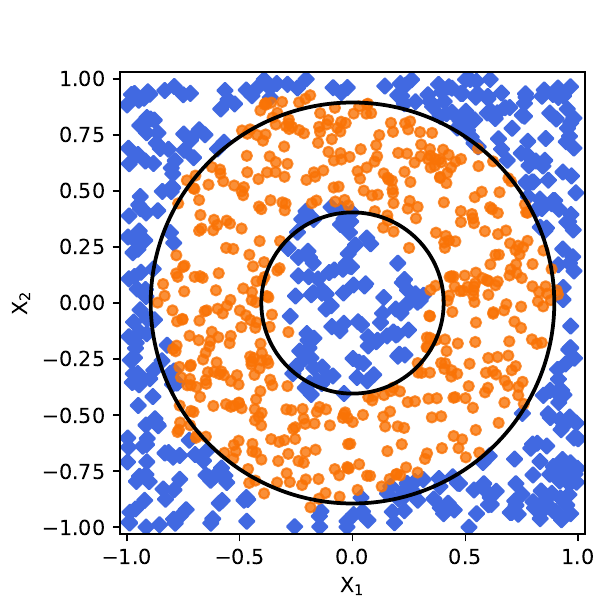}
    }
    \caption{Classifier test results for non-convex and non-trivial topology problems. The ion trap based QPU classifier performed on $1000$ random test data points is depicted in blue for points within and orange outside the boundary separating the circular feature shown in solid line. The results {\sl non-convex} and {\sl crown} datasets are computed using  4 layers, both from QPU and simulation. Notice that the border between classes in the experimental results is not as sharply defined as in the simulated classification. This difference is due to the uncertainty of the quantum measurements.
}
   \label{fig:nonconvex_crown}
    
         \begin{center}
        \subfigure[\hspace{1mm} {\sl hypersphere} - QPU]{
            \label{fig:hypersphere_qpu}
            \includegraphics[width=0.28\textwidth]{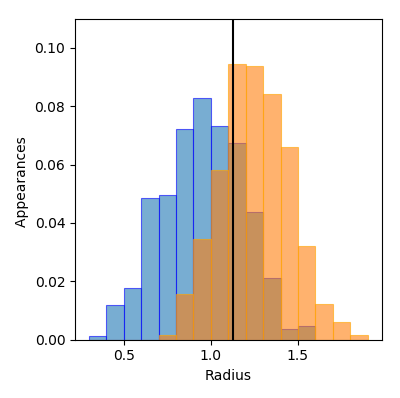}
        }
        \subfigure[\hspace{1mm} {\sl hypersphere} -  Simulation]{
           \label{fig:hypersphere_sim}
           \includegraphics[width=0.28\textwidth]{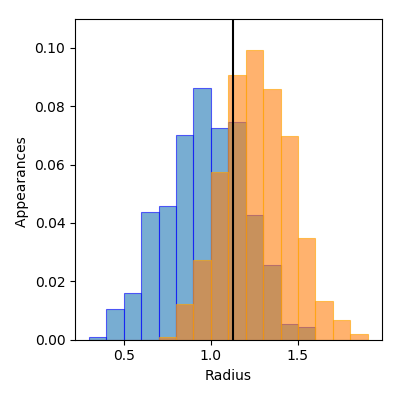}
        }
        \subfigure[\hspace{1mm} Experimental optimization]{
            \label{fig:hypersphereerror}
            \includegraphics[width=0.35\textwidth]{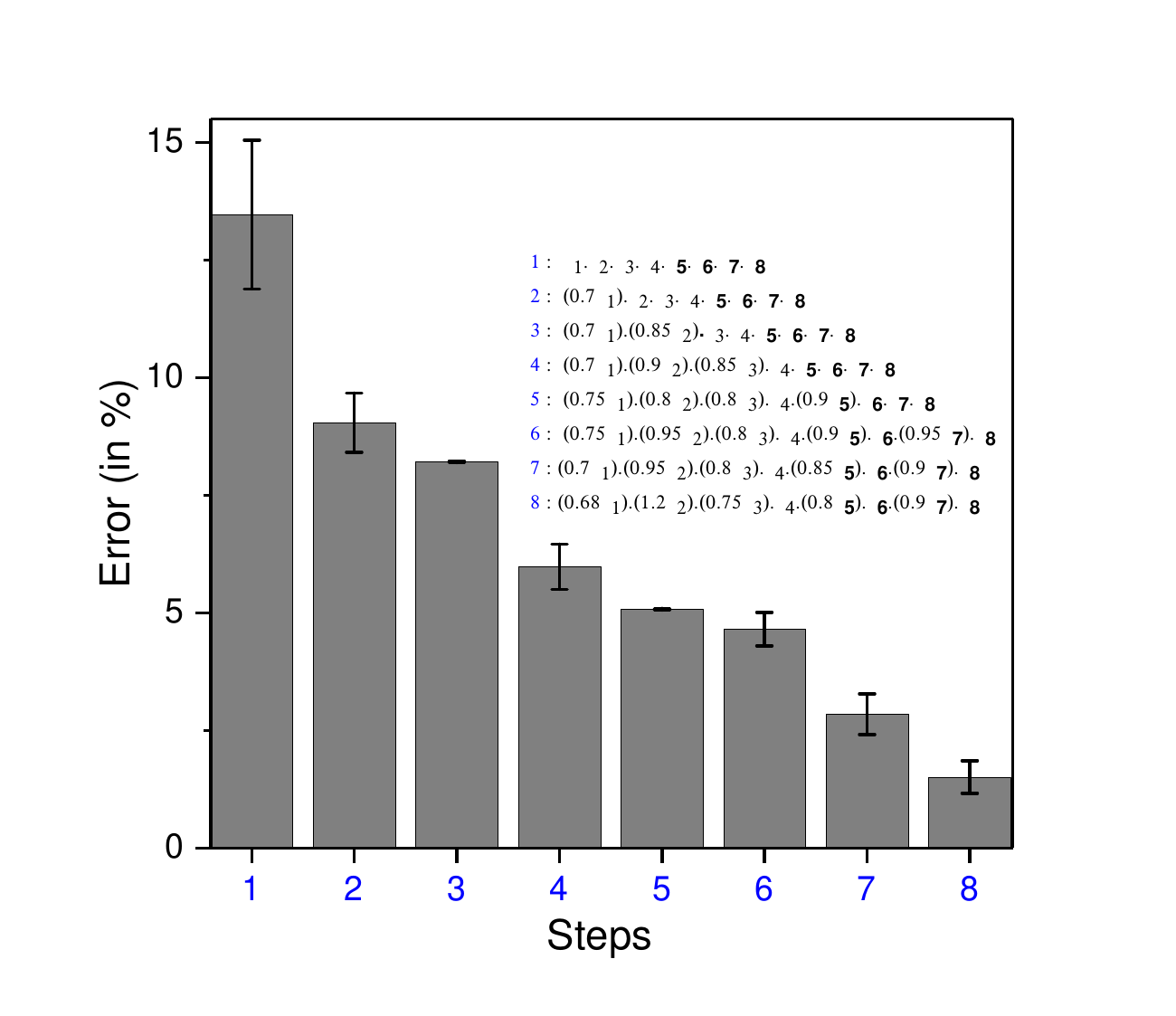}
        }
    \end{center}
 \caption{Binary classification of the {\sl hypersphere} dataset. The histograms in (a) and (b) represent the class association of points within a hyper shell (given by the bin-width) denoted by blue (classified as within the hyper-sphere) and orange (outside the hyper-sphere) for QPU (a) and simulation (b). The overlap region shows the ambiguity in classifying the points within a certain hyper radius.  The accuracy of the QPU is improved by performing the experimental optimization near the vicinity of the simulated optima in a series of ten training steps. The reduction in the error with respect to the simulated results (b) is shown in (c).}
   \label{fig:hyper_4d}

    \subfigure[\hspace{1mm} {\sl tricrown} - QPU\label{fig:tricrown_QPU}]{
    \includegraphics[width=.23\linewidth]{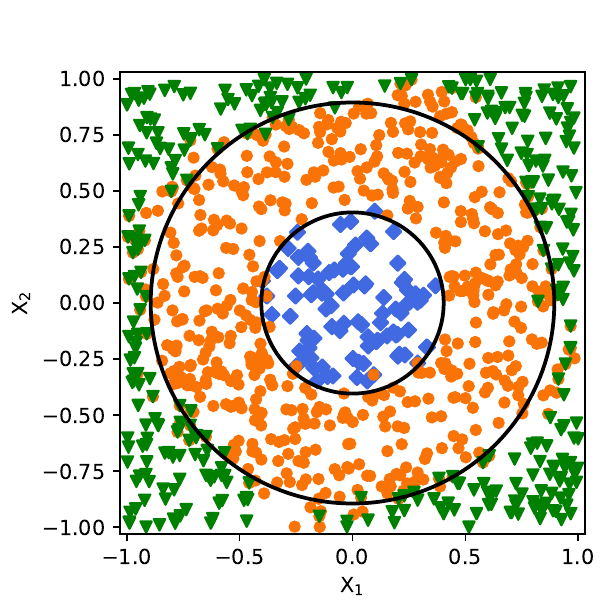}}
    \subfigure[\hspace{1mm} {\sl tricrown} - Simulation\label{fig:tricrown_sim}]{
    \includegraphics[width=.23\linewidth]{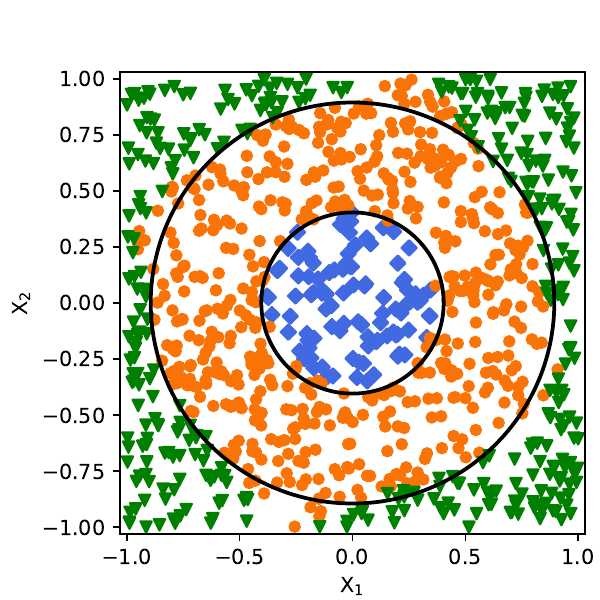}
    }
       \caption{Multi-class classification for the 3-class {\sl tricrown} problem on the QPU (left) and the simulation (right). Results include $1000$ random data points, the classes are depicted in different colors and symbols, and the boundaries are defined by solid black lines. Notice that the border between classes in the experimental results is not as sharply defined as in the simulated classification. This difference is due to the uncertainty of the quantum measurements.
}
    \label{fig:tricrown}
\end{figure*}

The results depicted in Fig.~\ref{fig:nonconvex_crown} correspond to additional binary classification problems. In both cases, classes are defined in such a way that each one fills half the total feature space. Notice that these classification problems are harder than the circle discussed above~\cite{Class_complexity}. The non-convexity feature of the {\sl non-convex} dateset presents a non-trivial challenge. In the {\sl crown} case, the different classes are not only non-convex but also the in-out class is disjoint. In order to solve the problem, the mapping performed by the circuit must be able to reflect this property. Even though, the final result shows that all problems can be solved in the ion-trap QPU using the re-uploading scheme. The final accuracies obtained ($\mathcal{A}^{\rm sim}$) / ($\mathcal{A}^{*}$) are {\sl non-convex}:($92\%$) / ($95\%$); {\sl crown}: ($87\%$) / ($92\%$). 

For the {\sl tricrown} example depicted in Fig.~\ref{fig:tricrown}, the difficulty is to add a third class to the {\sl crown} problem. This, however, makes all regions in the feature space joint, but a non-trivial mapping is needed to transform the topology of the dataset to the target states in the Bloch sphere~\cite{perez-salinas_data_2020}. 

For Fig.~\ref{fig:hyper_4d} the classification results of the hypersphere are depicted both for QPU and simulation. In this case, only the radius and not the full description of the point is depicted for graphical representation. Notice that the boundary lies partially outside from the feature space, $r > 1$. This is due to geometrical reasons. For 4 and more dimensions, the n-ball with volume half of the $[-1, 1]^n$ hypercube does not fit into the hypercube itself. This changes the accuracy of the random classifier, but it is not taken into account in this analysis. 

\end{document}